\documentclass[journal,twoside,web]{ieeecolor}

\usepackage{lcsys}

\usepackage{booktabs}
\usepackage{adjustbox}

\usepackage[top=20.2mm,bottom=18mm,left=16.9mm,right=20.2mm]{geometry}

\usepackage{wrapfig}
\usepackage{lipsum}
\usepackage{amssymb}
\usepackage{amsmath}
\usepackage{siunitx}
\usepackage{mathtools}
\usepackage{hyperref}
\let\labelindent\relax
\usepackage[inline]{enumitem}
\usepackage[ruled,vlined,linesnumbered]{algorithm2e}
\usepackage[noend]{algpseudocode}
\usepackage{ifthen}
\usepackage{dsfont}
\usepackage[backend=biber,style=numeric-comp,sorting=none,maxbibnames=99]{biblatex}
\usepackage{balance}
\usepackage{dirtytalk}
\usepackage{afterpage}
\usepackage{bm}

\usepackage{times}

\allowdisplaybreaks

\newcommand{\A}{\mathcal{A}}
\newcommand{\Avoid}{\mathbb{A}}
\newcommand{\B}{\mathcal{B}}
\newcommand{\C}{\mathcal{C}}
\newcommand{\D}{\mathcal{D}}

\newcommand{\F}{\mathcal{F}}
\newcommand{\G}{\mathcal{G}}

\newcommand{\K}{\mathcal{K}}
\newcommand{\KKT}{\text{KKT}}

\newcommand{\N}{\mathbb{N}}

\newcommand{\ra}{\rightarrow}
\newcommand{\R}{\mathbb{R}}
\newcommand{\Ra}{\Rightarrow}

\newcommand{\Safe}{\mathbb{S}}

\newcommand{\urk}{{\urcorner{k}}}
\newcommand{\urs}{{\urcorner{s}}}

\newcommand{\Z}{\mathcal{Z}}

\newcommand{\st}{\text{ s.t. }}
\newcommand{\paren}[1]{{({#1})}}

\newcommand{\boldlambda}{\boldsymbol{\lambda}}
\newcommand{\boldnu}{\boldsymbol{\nu}}
\newcommand{\bolddelta}{\boldsymbol{\delta}}
\newcommand{\boldeta}{\boldsymbol{\eta}}

\newcommand{\boldg}{\textbf{g}}
\newcommand{\boldh}{\textbf{h}}

\newcommand{\boldx}{\textbf{x}}
\newcommand{\boldu}{\textbf{u}}
\newcommand{\boldw}{\textbf{w}}
\newcommand{\boldy}{\textbf{y}}
\newcommand{\bolde}{\textbf{e}}
\newcommand{\boldv}{\textbf{v}}
\newcommand{\boldz}{\textbf{z}}
\newcommand{\boldzero}{\textbf{0}}

\newcommand{\boldU}{\textbf{U}}
\newcommand{\boldY}{\textbf{Y}}

\newtheorem{remark}{Remark}
\newtheorem{assumption}{Assumption}
\newtheorem{proposition}{Proposition}

\newtheorem{lemma}{Lemma}
\newtheorem{theorem}{Theorem}

\newcommand{\revisionRR}[1]{\textcolor[rgb]{0,0,0}{#1}}

\newcommand{\revision}[1]{\textcolor[rgb]{0,0,0}{#1}}

\newcommand{\revisionA}[1]{\textcolor[rgb]{0,0,0}{#1}}

\newcommand{\revisionB}[1]{\textcolor[rgb]{0,0,0}{#1}}

\bibliography{references}

\title{
Learning Constraints from Stochastic Partially-Observed Closed-Loop Demonstrations
}

\author{Chih-Yuan Chiu$^{1\star}$, \IEEEmembership{Member, IEEE}, Zhouyu Zhang$^{1\star}$, and Glen Chou$^2$, \IEEEmembership{Member, IEEE}
\thanks{Georgia Institute of Technology, Atlanta, GA, USA. $^{1}$School of Electrical and Computer Engineering (\texttt{\{cyc, zzhang3097\} at gatech dot edu}). $^{\star}$Equal contribution. $^{2}$Schools of Cybersecurity and Privacy and of Aerospace Engineering (\texttt{chou at gatech dot edu}).}
}

\def\BibTeX{{\rm B\kern-.05em{\sc i\kern-.025em b}\kern-.08em
T\kern-.1667em\lower.7ex\hbox{E}\kern-.125emX}}

\def\BibTeX{{\rm B\kern-.05em{\sc i\kern-.025em b}\kern-.08em
T\kern-.1667em\lower.7ex\hbox{E}\kern-.125emX}}

\begin{document}

\maketitle

\thispagestyle{empty}
\pagestyle{empty}




\begin{abstract}
We present a method for learning unknown parametric constraints from locally-optimal input-output trajectory data. 
We assume the data is generated by rollouts of stochastic nonlinear dynamics, 
\revision{under a single state or output feedback law and initial condition but distinct noise realizations}, 
to robustly satisfy underlying constraints despite worst-case noise outcomes.
We encode the Karush-Kuhn-Tucker (KKT) conditions of this robust optimal feedback control problem within a feasibility problem to recover constraints consistent with the local optimality of the demonstrations. We prove that our constraint learning method (i) accurately recovers the demonstrator's policy, and (ii) conservatively estimates the set of policies that ensure constraint satisfaction despite worst-case noise realizations. Moreover, we perform sensitivity analysis, proving that when demonstrations are corrupted by transmission error, the inaccuracy in the learned feedback law scales linearly in the error magnitude. \revisionRR{Empirically,} our method accurately recovers unknown constraints from simulated noisy, closed-loop demonstrations generated using dynamics, both linear and nonlinear, (e.g., unicycle and quadrotor) and a range of feedback mechanisms.
\end{abstract}


\section{Introduction}
\label{sec: Introduction}

\IEEEPARstart{I}{n} safety-critical robotics, autonomous agents must infer and enforce hard constraints to guarantee safe operation, such as for collision avoidance in crowded environments. 
Learning from demonstration (LfD) enables robots to recover unknown parametric constraints from trajectory data 
\cite{Chou2020LearningConstraintsFromLocallyOptimalDemonstrationsUnderCostFunctionUncertainty, Menner2021ConstrainedInverseOptimalControlWithApplicationtoaHumanManipulationTask, Armesto2017EfficientLearningofConstraintsandGenericNullSpacePolicies}.
The key idea is that locally-optimal demonstrations generated from a constrained optimal control problem satisfy an associated set of Karush-Kuhn-Tucker (KKT) conditions, which can be inverted to recover constraints via inverse optimal control (IOC). 
Prior work has shown this is possible for \textit{open-loop} state and control demonstrations under \textit{deterministic} dynamics \cite{Chou2020LearningConstraintsFromLocallyOptimalDemonstrationsUnderCostFunctionUncertainty}.
In practice, however, agents operate with state uncertainty and stochastic dynamics. 
Thus, instead of open-loop plans, a demonstrator must apply an \textit{output feedback control policy} to \textit{partial, noisy observations}. 
Existing methods cannot recover constraints from such input-output data.

To address this gap, we make the following contributions.
We present a novel method
to infer \textit{a priori} unknown parametric constraints from input-output data generated under 
linear time-varying (LTV) 
stochastic dynamics with partial observations and output noise. 
Our method extends the IOC-based constraint learning method in \cite{Chou2020LearningConstraintsFromLocallyOptimalDemonstrationsUnderCostFunctionUncertainty} to the partial and noisy observation setting (Sec. \ref{sec: Methods}). We evaluate our method on linear and nonlinear (e.g., unicycle, quadrotor) dynamics and a variety of feedback controllers, successfully recovering constraints from the generated demonstrations (Sec. \ref{sec: Experiments}).

Our work draws from and contributes to IOC-based LfD for cost and constraint inference. Specifically,  \cite{englert2017inverse, Menner2021ConstrainedInverseOptimalControlWithApplicationtoaHumanManipulationTask} introduce IOC methods assuming known constraints, 
while 
\cite{ Zhang2024InverseOptimalControlforAveragedCost, Rickenbach2025InverseOptimalControlWithConstraintRelaxation, Molloy2018FiniteHorizonInverseOptimalControl, Asl2025DataDrivenIOC}
(resp., \cite{Chou2020LearningConstraintsFromLocallyOptimalDemonstrationsUnderCostFunctionUncertainty, McPherson2021MLConstraintInferenceFromStochasticDemonstrations}) enable cost (resp., constraint) inference via IOC.
Unlike \cite{McPherson2021MLConstraintInferenceFromStochasticDemonstrations}, which addresses stochastic demonstrations in discrete spaces, our method recovers constraints with provable accuracy guarantees in continuous space. 
Also, while 
\cite{Chou2020LearningConstraintsFromLocallyOptimalDemonstrationsUnderCostFunctionUncertainty}
use deterministic trajectories, we learn constraints from demonstrations generated via output feedback with noisy, partial observations.
On a technical level, we leverage system-level synthesis (SLS) \cite{anderson2019levelsynthesis, Leeman2023PredictiveSafetyFilterusingSystemLevelSynthesis, Leeman2025RobustNonlinearOptimalControlviaSLS} to encode output feedback policies for input-output demonstrations. Whereas prior SLS work focuses on controller design, we apply SLS to infer constraints from noisy, partially observed data. 

\textit{Notation}:
$\forall x_0, \cdots, x_T \in \R^n$, let $x_{t_1:t_2} := (x_{t_1}, \cdots, x_{t_2}) \in \R^{n(t_2 - t_1 + 1)}$
$\forall t_1, t_2 \in \{0,\ldots,T\} := [0, T]$ with $t_1 \leq t_2$. 
For $M_1, \cdots, M_K \in \R^{m \times n}$, let $M := \text{diag}\{M_1, \cdots, M_K\} \in \R^{mK \times nK}$ denote the block diagonal matrix with $M_k$ as the $k$-th block, $\forall k \in \{1, \ldots, K\} := [K]$. 
\revisionRR{We denote the $m \times m$ identity matrix, $m \times n$ zero matrix, and $n \times 1$ zero vector by $I_m$, $O_{m \times n}$, and $\textbf{0}_n$, respectively, $\forall m, n \in \N$,}
with subscripts omitted for dimensions clear from context.
\section{Problem Formulation}
\label{sec: Problem Formulation}

We formulate the robust optimal control problem used to generate demonstration data (Sec. \ref{subsec: Robust Optimal Control Problem Formulation}), present system level synthesis (SLS) for output feedback (Sec. \ref{subsec: System Level Synthesis (SLS) for Output Feedback}), discuss data generation (Sec. \ref{subsec: Demonstration Generation Process}), and state our problem (Sec. \ref{subsec: Problem Statement}).

\subsection{Robust Optimal Control Problem Formulation}
\label{subsec: Robust Optimal Control Problem Formulation}

Consider a stochastic dynamical system $f_t: \R^n \times \R^{n_i} \ra \R^n$ with additive dynamics and output noise, and initial condition fixed at $x_0 = \bar x$
as shown below:\footnote{Uncertainty in $x_0$ can be modeled 
\revisionRR{as a non-zero first component}
of the dynamics noise\revisionRR{, replacing $\boldzero_n$} (see \cite{Leeman2023PredictiveSafetyFilterusingSystemLevelSynthesis}); we do so in our experiments.}
\begin{subequations} \label{Eqn: Forward Dynamics and Output Maps}
\begin{alignat}{2} \label{Eqn: Forward Dynamics}
    x_{t+1} &= f_t(x_t, u_t) + w_t, \hspace{5mm} &&\forall t \in [0,T-1], \\ \label{Eqn: Forward Output Map}
    y_t &= C_t x_t + e_t, &&\forall t \in [0,T].
\end{alignat}
\end{subequations}
Above, $x_t \in \R^n$, $u_t \in \R^{n_i}$, $w_t \in \R^n$, $y_t \in \R^{n_o}$, and $e_t \in \R^{n_o}$ respectively denote 
the state, input, dynamics noise, output, and output noise at time $t \in [0, T]$.
Let $\boldx := x_{0:T} \in \R^{n(T+1)}$, \revisionRR{$\boldu := u_{0:T} \in \R^{n_i (T+1)}$, $\boldw := (\boldzero_n, w_{0:T-1}) \in \R^{n(T+1)}$, and $\boldy := y_{0:T} \in \R^{n_o(T+1)}, \bolde := e_{0:T} \in \R^{n_o (T+1)}$}.
We assume there exist known, compact sets $W \subset \R^n$ and $E \subset \R^{n_o}$ such that $w_t \in W \ \forall t \in [T-1]$ and $e_t \in E \ \forall t \in [T]$. Note that $W$ and $E$ can be estimated from data, e.g., via \cite{KnuthChou2023StatisticalSafetyandRobustnessGuaranteesforFeedbackMotionPlanningofUnknownUnderactuatedStochasticSystems}.

To ease exposition, unless noted otherwise, we assume linear time-varying (LTV) dynamics, i.e., $f_t(x_t, u_t) = A_t x_t + B_t u_t$. We add clarifying remarks when a result holds for LTV but not for general nonlinear dynamics
(Remarks \ref{Remark: Nonlinear dynamics, Output feedback error laws}, \ref{Remark: Nonlinear dynamics, Robust reformulation}, \ref{Remark: Nonlinear dynamics, nominal trajectory and control recovery via nonlinear regression}).



Let \revisionRR{$S_k$, $S_\urk(\theta^\star) \subset \revisionRR{\R^{(n+n_i)(T+1)}}$} respectively denote a \textit{known} and an \textit{unknown, parameterized} constraint set imposed on the state-control trajectory $(\boldx, \boldu)$, where $\theta^\star \in \R^d$ denotes a 
ground truth parameter vector \textit{a priori} unknown to the constraint learner. Concretely, \revisionRR{$S_k$} and \revisionRR{$S_\urk(\theta^\star)$} are respectively defined by inequality constraints $g_{\beta,k}: \revisionRR{\R^{(n+n_i)(T+1)}} \ra \revisionRR{\R}, \forall \beta \in [n_k]$, and $g_{\beta,\urk}: \revisionRR{\R^{(n+n_i)(T+1)}} \times \R^d \ra \revisionRR{\R}, \forall \beta \in [n_\urk]$, as shown below:
\begin{subequations} 
\begin{align} 
\nonumber
    \revisionRR{S_k} &:= \{(\boldx, \boldu): g_{\beta,k}(\boldx, \boldu) \leq 0, \forall \beta \in [n_k]
    \}, \\ 
    \nonumber
    \revisionRR{S_\urk}(\theta^\star) &:= \{ (\boldx, \boldu): g_{\beta,\urk}(\boldx, \boldu, \theta^\star) \leq 0, \forall \beta \in [n_\urk] 
    \},
\end{align}
\end{subequations}

For each $t \in [0, T-1]$, let $\pi_t: \R^{n_o(t+1)} \ra \R^{n_i}$ be a causal output feedback controller which maps realized outputs \revisionB{$y_{0:t}$} to controls $u_t$, and let $\pi(\boldy) := (\pi_0(y_0), \cdots, \pi_{T-1}(\revisionRR{y_{0:T}}))$.
We define $\Pi := \Pi(W, E, \revisionRR{S_k}, \revisionRR{S_\urk}, \theta^\star)$ to be the set of all causal output feedback laws $\pi(\cdot)$ which generate trajectories that robustly satisfy the constraint sets \revisionRR{$S_k$} and \revisionRR{$S_\urk(\theta^\star)$} despite worst-case noise realizations in $W$ and $E$.
We then define a cost map $J: \revisionRR{\R^{(n+n_i)(T+1)}} \ra \R$ and formulate the following robust control problem, which is generally computationally intractable due to the optimization over 
$\pi(\cdot)$:
\begin{subequations} \label{Eqn: Original, Robust Forward}
\begin{align} \label{Eqn: Original, Robust Forward, Objective}
    \min_{\pi(\cdot)}. \max_{\boldw, \bolde, \boldx, \boldu, \boldy} \hspace{3mm} &J(\boldx, \boldu) \\ \label{Eqn: Original, Robust Forward, Constraint 1, Dynamics}
    &(\boldw, \bolde, \boldx, \boldu, \boldy) \text{ satisfy } \eqref{Eqn: Forward Dynamics and Output Maps}, \ x_0 = \bar x \\ \label{Eqn: Original, Robust Forward, Constraint 2, Noise Terms}
    &\boldw \in W^{T+1}, \bolde \in E^{T+1}, \\ \label{Eqn: Original, Robust Forward, Constraint 3, Policy}
    &\pi(\cdot) \in \Pi(W, E, \revisionRR{S_k}, \revisionRR{S_\urk}, \theta^\star)\revisionB{.}
\end{align}
\end{subequations}

\subsection{System Level Synthesis (SLS) for Output Feedback}
\label{subsec: System Level Synthesis (SLS) for Output Feedback}

A common \revisionB{approach used rather than} solving \eqref{Eqn: Original, Robust Forward} is to find (i) nominal state-control trajectories $(\boldz, \boldv) \in \revisionRR{\R^{(n+n_i)(T+1)}}$ that minimize cost in the noise-free case and (ii) linear output feedback controllers to reject noise \cite{ZhouTzoumas2023SafeControlofPartiallyObservedLTVSystems, Leeman2023PredictiveSafetyFilterusingSystemLevelSynthesis, ManchesterSlotine2017ControlContractionMetrics}. Equivalently, we assume each 
feedback law $\pi_t(\cdot)$ for $t \in [0, \revisionRR{T}]$ has the form:
\begin{align} \label{Eqn: Output Error Feedback, time t}
    u_t = \pi_t(y_{0:t}) &= v_t + \sum_{\tau=0}^t K_{t, \tau} (y_\tau - C_\tau z_\tau),
\end{align}
where $K_{t, \tau} \in \R^{n_i \times n_o}$, $\forall t \in \revision{[0, T]}$, $\tau \in [0, t]$.
To write \eqref{Eqn: Output Error Feedback, time t} compactly, let $\A := \text{diag}\{A_0, \cdots, A_{T-1}, \revisionRR{O_{n \times n}} \} \in \R^{n\revisionRR{(T+1)} \times n\revisionRR{(T+1)}}$, $\B := \text{diag}\{B_0, \cdots, B_{T-1}, \revisionRR{O_{n \times n_i}}\} \in \R^{n\revisionRR{(T+1)} \times n_i \revisionRR{(T+1)}}$, $\C := \text{diag}\{C_0, \cdots, C_{T-1}, \revisionRR{C_T}\} \in \R^{n_o \revisionRR{(T+1)} \times n\revisionRR{(T+1)}}$, and define $\K \in \R^{n_i \revisionRR{(T+1)} \times n_o \revisionRR{(T+1)}}$ 
by:
\begin{align} \label{Eqn: Output Feedback Matrix, K, Lower Triangular}
    \K &:= \begin{bmatrix}
        K_{0, 0} & O & \cdots & O \\[-7pt]
        K_{1, 0} & K_{1, 1} & \ddots & O \\[-3pt]
        \vdots & \vdots & \ddots & \vdots \\[-7pt]
        \revisionRR{K_{T, 0}} & \revisionRR{K_{T, 1}} & \ddots & \revisionRR{K_{T, T}}
    \end{bmatrix}
\end{align}
Then \eqref{Eqn: Output Error Feedback, time t} can be written more compactly as:
\begin{align} \label{Eqn: (Compact) Output Error Feedback, time t}
    \boldu = \pi(y_{0:T}) &= \boldv + \K (\boldy - \C \boldz).
\end{align}
The policy optimization in \eqref{Eqn: Original, Robust Forward} can now be recast over the nominal state-control trajectory $(\boldz, \boldv)$ and the output error feedback gain $\K$ in \eqref{Eqn: Output Feedback Matrix, K, Lower Triangular}. These two optimizations may be decoupled, as in methods based on control contraction metrics (CCM) \cite{ManchesterSlotine2017ControlContractionMetrics}, or coupled, as in system-level synthesis (SLS)-based robust control designs \cite{ZhouTzoumas2023SafeControlofPartiallyObservedLTVSystems, Leeman2023PredictiveSafetyFilterusingSystemLevelSynthesis}. 

Output feedback SLS \cite{ZhouTzoumas2023SafeControlofPartiallyObservedLTVSystems, anderson2019levelsynthesis} parameterizes the feedback $\K$ via a\textit{
system response} $\Phi$, enabling joint optimization over $\K$ and $(\boldz, \boldv)$.
Concretely, define the state, control, and output error respectively by $\Delta \boldx := \boldx - \boldz$, $\Delta \boldu := \boldu - \boldv$, and $\Delta \boldy := \boldy - \C \boldz$. Since $(\boldz, \boldv)$ satisfy the noise-free version of \eqref{Eqn: Forward Dynamics and Output Maps}, i.e., $z_{t+1} = A_t z_t + B_t v_t$ for all $t \in [T]$, the error dynamics are:
\begin{subequations} \label{Eqn: Error Dynamics and Output Maps}
\begin{alignat}{2} \label{Eqn: Error Dynamics}
    \Delta x_{t+1} &= A_t \Delta x_t + B_t \Delta u_t + w_t, \hspace{5mm} &&\forall t \in [0,T-1], \\ 
    \label{Eqn: Error Input Map}
    \Delta u_t &= \sum_{\tau=0}^t K_{t,\tau} (C_\tau \Delta x_\tau + e_\tau), &&\forall t \in [0, \revisionRR{T}],
\end{alignat}
\end{subequations}
with $\Delta x_0 = 0$. We concatenate \eqref{Eqn: Error Dynamics and Output Maps} across times $t \in [T]$ by introducing the downshift operator $\Z$, given by:
\begin{align} \label{Eqn: Z, Downshift Operator}
    \Z &:= \begin{bmatrix}
        O_{n \times n\revisionRR{T}} & O_n \\
        I_{n\revisionRR{T}} & O_{n\revisionRR{T} \times n} 
    \end{bmatrix} \in \revisionRR{\R^{n(T+1) \times n(T+1)}}.
\end{align}
This enables us to rewrite \eqref{Eqn: Error Dynamics and Output Maps} as:
\begin{subequations} \label{Eqn: Error Dynamics and Output Maps, stacked}
\begin{align} \label{Eqn: Error Dynamics, stacked}
    \Delta \boldx &= \Z \A \Delta \boldx + \Z \B \Delta \boldu + \boldw, \\ \label{Eqn: Error Output, stacked}
    \Delta \boldu &= \K (\C \Delta \boldx + \bolde).
\end{align}
\end{subequations}
Rearranging terms, 
we can rewrite \eqref{Eqn: Error Dynamics and Output Maps, stacked} as:
\begin{align}
    \begin{bmatrix}
        \Delta \boldx \\ \Delta \boldu
    \end{bmatrix}
    &= \begin{bmatrix}
        \Phi_{xw} & \Phi_{xe} \\
        \Phi_{uw} & \Phi_{ue}
    \end{bmatrix}
    \begin{bmatrix}
        \boldw \\ \bolde
    \end{bmatrix} := \Phi \begin{bmatrix}
        \boldw \\ \bolde
    \end{bmatrix},
\end{align}

\noindent
where the \textit{system response} $\Phi \in \R^{\revisionRR{((n+n_i)(T+1)) \times (n+n_o) (T+1)}}$ is block-wise defined from $\Z, \A, \B, \C$, and $\K$, as follows:
\begin{subequations} \label{Eqn: Phi from K}
\begin{align} \label{Eqn: Phi xw from K}
    \Phi_{xw} &:= (I - \Z \A - \Z \B \K \C)^{-1} \\ \label{Eqn: Phi xe from K}
    \Phi_{xe} &:= (I - \Z \A - \Z \B \K \C)^{-1} \Z \B \K \\ \label{Eqn: Phi uw from K}
    \Phi_{uw} &:= \K \C (I - \Z \A - \Z \B \K \C)^{-1} \\ \label{Eqn: Phi ue from K}
    \Phi_{ue} &:= \K \C (I - \Z \A - \Z \B \K \C)^{-1} \Z \B \K + \K.
\end{align}
\end{subequations}
Conversely, we also observe that:
\begin{align} \label{Eqn: K from Phi}
    \K &= \Phi_{ue} - \Phi_{uw} \Phi_{xw}^{-1} \Phi_{xe}.
\end{align}

Note that \cite{ZhouTzoumas2023SafeControlofPartiallyObservedLTVSystems}
proves that the closed-loop system response $\Phi$ characterizes all causal output feedback laws $\K$:

\begin{proposition}\cite[Prop. 1]{ZhouTzoumas2023SafeControlofPartiallyObservedLTVSystems}
If there exists a lower block triangular matrix $\K$ satisfying \eqref{Eqn: Error Dynamics and Output Maps, stacked}, then $\Phi_{xw}$, $\Phi_{xe}$, $\Phi_{uw}$, and $\Phi_{ue}$, as computed by \eqref{Eqn: Phi from K}
satisfy the affine equalities:
\begin{subequations} \label{Eqn: Phi, Affine Constraints}
\begin{align} \label{Eqn: Phi, Affine Constraints, Controllability}
    \begin{bmatrix}
        I - \revisionB{\mathcal{Z}} \A & - \revisionB{\mathcal{Z}} \B
    \end{bmatrix} \Phi &= 
    \begin{bmatrix}
        I & O
    \end{bmatrix}, \\ \label{Eqn: Phi, Affine Constraints, Observability}
    \Phi \begin{bmatrix}
        I - \revisionB{\mathcal{Z}} \A \\ -\C
    \end{bmatrix} &= 
    \begin{bmatrix}
        I \\ O
    \end{bmatrix}, \\ \label{Eqn: Phi, Affine Constraints, Upper Triangular}
    \Phi_{xw}, \Phi_{xe}, \Phi_{uw}, \Phi_{ue} &\text{ lower block triangular as in \eqref{Eqn: Output Feedback Matrix, K, Lower Triangular}}
\end{align}
\end{subequations}
Conversely, if $\Phi_{xw}$, $\Phi_{xe}$, $\Phi_{uw}$, and $\Phi_{ue}$ 
satisfy \eqref{Eqn: Phi, Affine Constraints}, then $\K$, as computed by \eqref{Eqn: K from Phi}, is lower block triangular.
\end{proposition}


\begin{remark} \label{Remark: Nonlinear dynamics, Output feedback error laws}
If the dynamics \eqref{Eqn: Forward Dynamics} are nonlinear, then \eqref{Eqn: Phi, Affine Constraints}, with each $A_t$, $B_t$ replaced by the Jacobian linearization of $f_t$
about given linearization points,
would parameterize a subset of all causal affine output error feedback laws.
\end{remark}

Using \cite[Prop. 1]{ZhouTzoumas2023SafeControlofPartiallyObservedLTVSystems}, we 
\revisionA{modify}
\eqref{Eqn: Original, Robust Forward} as an optimization problem over nominal trajectories $(\boldz, \boldv)$ and system responses $\Phi$. Concretely, given 
$(\boldz, \boldv)$ 
and $\Phi$, 
we characterize the worst-case
constraint realizations via the functions 
$\tilde g_{\beta,k}(\cdot)$, $\forall \beta \in [n_k]$, and $\tilde g_{\beta,\urk}(\cdot)$, $\forall \beta \in [n_\urk]$, defined as:
\begin{align} 
    \nonumber
    \tilde g_{\beta, k}(\boldz, \boldv, \Phi) &:= \max_{\substack{\boldw \in W^{T+1} \\ \bolde \in E^{T+1}}} g_{\beta, k} \left( \begin{bmatrix}
        \boldz \\ \boldv
    \end{bmatrix} + \Phi \begin{bmatrix}
        \boldw \\ \bolde
    \end{bmatrix} \right), \\ \nonumber  
    \tilde g_{\beta, \urk}(\boldz, \boldv, \Phi, \theta^\star) &:= \max_{\substack{\boldw \in W^{T+1} \\ \bolde \in E^{T+1}}} g_{\beta, \urk}\left( \begin{bmatrix}
        \boldz \\ \boldv
    \end{bmatrix} + \Phi \begin{bmatrix}
        \boldw \\ \bolde
    \end{bmatrix}, \theta^\star \right).
\end{align}
We 
stack constraints $\tilde g_{\beta,k}$ (resp., $\tilde g_{\beta,\urk}$) across indices $\beta$ to form the vector-valued maps $\tilde \boldg_k$ (resp., $\tilde \boldg_\urk$), and
define $\tilde \boldh$ to encode 
\eqref{Eqn: Phi, Affine Constraints} and
the dynamics $z_{t+1} = A_t z_t + B_t v_t \ \forall t \in [T]$, 
as $\tilde \boldh(\cdot) = \revisionRR{\boldzero}$.
We then pose the optimization program \eqref{Eqn: Reformulated, Robust Forward}, which presents a feasible solution to 
\revisionA{\eqref{Eqn: Original, Robust Forward, Constraint 1, Dynamics}-\eqref{Eqn: Original, Robust Forward, Constraint 3, Policy}}
\cite[Prop. 4]{Leeman2023PredictiveSafetyFilterusingSystemLevelSynthesis}:
\begin{subequations} \label{Eqn: Reformulated, Robust Forward}
\begin{align} \label{Eqn: Reformulated, Robust Forward, Objective}
    \min_{\boldz, \boldv, \Phi}
    \hspace{5mm} 
    &\revision{J(\boldz, \boldv)} \\ 
    \label{Eqn: Reformulated, Robust Forward, Equality Constraints}
    &\tilde \boldh(\boldz, \boldv, \Phi) = \boldzero, \\ \label{Eqn: Reformulated, Robust Forward, Inequality Constraints, Known}
    &\tilde \boldg_k(\boldz, \boldv, \Phi) \leq \boldzero, \\ \label{Eqn: Reformulated, Robust Forward, Inequality Constraints, Unknown}
    &\tilde \boldg_\urk(\boldz, \boldv, \Phi, \theta^\star) \leq \boldzero.
\end{align}
\end{subequations}

\begin{remark} \label{Remark: Nonlinear dynamics, Robust reformulation}
If the dynamics \eqref{Eqn: Forward Dynamics} are nonlinear, 
$\tilde{\boldh}$ must 
be defined to 
encode $z_{t+1} = f_t(z_t, u_t), \forall t \in [0,T]$, as well as \eqref{Eqn: Phi, Affine Constraints}, with each $A_t, B_t$ replaced by the Jacobian linearization of $f_t$. Moreover,
we must define 
$\tilde{\textbf{g}}_k$ and $\tilde{\textbf{g}}_\urk$ 
while accounting for the
linearization error related to each $f_t$ \revisionRR{\cite[Sec. III]{Leeman2025RobustNonlinearOptimalControlviaSLS}}.
\end{remark}

\begin{remark}
For control schemes where an output feedback law $\bar \K$ 
is designed independently 
of the
optimal nominal trajectory (e.g., CCM methods), we can substitute $\K = \bar \K$ into \eqref{Eqn: Phi from K},
and append the resulting equalities (defining the resulting system response $\Phi$) to the constraint set of \eqref{Eqn: Reformulated, Robust Forward}.


\end{remark}

Given a parameter value $\theta \in \R^d$, let the set of corresponding safe (resp., unsafe) nominal trajectories and system responses, given below by $\revisionRR{\Safe}(\theta)$ (resp., $\revisionRR{\Avoid}(\theta)$), be defined by:
\begin{align}
    \revisionRR{\Safe}(\theta) &:= \{ (\boldz, \boldv, \Phi): \tilde \boldg_\urk(\boldz, \boldv, \Phi, \theta) \leq \boldzero \}, \\
    \revisionRR{\Avoid}(\theta) &:= \revisionRR{\Safe}(\theta)^c,
\end{align}
\revisionRR{where $(\cdot)^c$ indicates the complement of a set.}
In particular, $\revisionRR{\Safe}(\theta)$ contains the feasible set of \eqref{Eqn: Reformulated, Robust Forward}. 

\begin{remark}
$\tilde g_{\beta, k}(\cdot)$ and $\tilde g_{\beta, \urk}(\cdot)$ can be written in closed form for certain constraint functions $g_{\beta, k}(\cdot)$ and $g_{\beta, \urk}(\cdot)$ and disturbance sets $W$ and $E$ \cite[Prop. 4]{Leeman2023PredictiveSafetyFilterusingSystemLevelSynthesis}, \cite[Prop III.3]{Leeman2025RobustNonlinearOptimalControlviaSLS}.
\end{remark}

\subsection{Demonstration Generation Process}
\label{subsec: Demonstration Generation Process}

We describe how a set of $D$ input-output trajectory demonstrations
$\D := \{(\boldu^\paren{d}, \boldy^\paren{d}) \in \revisionRR{\R^{(n_i+n_o)(T+1)}}: d \in [D] \}$ are generated for constraint learning. First, the demonstrator solves \eqref{Eqn: Reformulated, Robust Forward} to compute a locally-optimal nominal trajectory $(z^\star, v^\star)$ and system response $\Phi^\star$, respectively, and computes the corresponding output feedback law $\K^\star$ by substituting $\Phi = \Phi^\star$ into \eqref{Eqn: K from Phi}. Then, $\forall d \in [D]$, the demonstrator
generates 
$(\boldu^\paren{d}, \boldy^\paren{d})$ 
by unrolling the dynamics \eqref{Eqn: Forward Dynamics}, and applying output map \eqref{Eqn: Forward Output Map} and output feedback law \eqref{Eqn: Output Error Feedback, time t}:
\begin{subequations} \label{Eqn: Demonstration Generation}
\begin{align} \label{Eqn: Demonstration Generation, Dynamics Map}
    x_{t+1}^\paren{d} &:= A_t x_t^\paren{d} + B_t u_t^\paren{d} + w_t^\paren{d} \\ \label{Eqn: Demonstration Generation, Output Map}
    y_t^\paren{d} &= C_t x_t^\paren{d} + e_t^\paren{d} \\ \label{Eqn: Demonstration Generation, Control Design via Output Feedback}
    u_t^\paren{d} &= v_t^\star + \textstyle \sum_{\tau=0}^t K_{t,\tau}^\star (y_\tau^\paren{d} - C z_\tau^\star).
\end{align}
\end{subequations}
Above, $(\boldw^\paren{d}, \bolde^\paren{d}) \in W^{T+1} \times E^{T+1}$ describes the dynamics and output noise realizations for the $d$-th input-output trajectory, for each $d \in [D]$. Finally, the constraint learner receives the \textit{perturbed} demonstration set $\tilde D := \{(\tilde{\boldu}^\paren{d}, \tilde{\boldy}^\paren{d}) \in \R^{n_i \revisionRR{(T+1)}} \times \R^{n_o \revisionRR{(T+1)}}: d \in [D] \}$, with:
\begin{equation} \label{Eqn: Transmission Error Perturbation}
    \tilde{\boldu}^\paren{d} = \boldu^\paren{d} + \bolddelta_u^\paren{d},\quad\quad \tilde{\boldy}^\paren{d} = \boldy^\paren{d} + \bolddelta_y^\paren{d},
\end{equation}
where $(\bolddelta_u^\paren{d}, \bolddelta_y^\paren{d}) \in \R^{n_i \revisionRR{(T+1)}} \times \R^{n_o \revisionRR{(T+1)}}$ represent possible \textit{transmission errors} that may corrupt the constraint learner's reception of the generated input-output demonstrations. 

\begin{remark}
Our 
methods also extend to learn constraints from 
input-output trajectories generated using 
\textit{multiple} nominal trajectories and feedback laws\revisionRR{, provided the learner is aware which input-output trajectories are identified with which nominal trajectory and feedback law}. 
\end{remark}


\subsection{Problem Statement}
\label{subsec: Problem Statement}

We aim
to infer the unknown constraint parameter $\theta^\star$ from the perturbed
demonstration set $\tilde D$ described 
above, 
system equations \eqref{Eqn: Forward Dynamics and Output Maps}, cost $J$,
and constraints $g_{\beta,k}$, $g_{\beta,\urk}$, $\tilde g_{\beta,k}$, and $\tilde g_{\beta,\urk}$.
We also aim to 
learn
nominal trajectories and output feedback laws that are guaranteed to lie within the safe set $\revisionRR{\Safe}(\theta)$
despite worst-case dynamics and output noise outcomes.

\section{Methods}
\label{sec: Methods}


\subsection{Recovering the Nominal Trajectory, Output Feedback, and Constraint Parameters}
\label{subsec: Recovering the Nominal Trajectory, Output Feedback, and Constraint Parameters}

First, we formulate a linear least-squares (LLS) program to infer the demonstrator's output feedback policy $\K^\star$ from the perturbed demonstration set $\tilde D$ received by the constraint learner. Concretely, from \eqref{Eqn: Demonstration Generation, Control Design via Output Feedback} and \eqref{Eqn: Transmission Error Perturbation}, we obtain:
\begin{align} 
    \label{Eqn: Demonstration Generation, Control Design via Output Feedback, Stacked}
    &\tilde{\boldu}^\paren{d} - \bolddelta_u^\paren{d} = \boldv^\star + \K^\star (\tilde{\boldy}^\paren{d} - \C \boldz^\star - \bolddelta_y^\paren{d}),
    \\ \label{Eqn: tilde u, K star, tilde y, affine equality}
    \Ra \ &(\tilde{\boldu}^\paren{d} - \tilde{\boldu}^\paren{d-1}) - (\bolddelta_u^\paren{d} - \bolddelta_u^\paren{d-1}) \\ \nonumber
    &\hspace{5mm} = \K^\star \big( (\tilde{\boldy}^\paren{d} - \tilde{\boldy}^\paren{d-1}) - (\bolddelta_y^\paren{d} - \bolddelta_y^\paren{d-1}) \big)
\end{align}
We define 
$\tilde{\boldU} \in \R^{n_i \revisionRR{(T+1)} \times (D-1)}$, $\tilde{\boldY} \in \R^{n_o \revisionRR{(T+1)} \times (D-1)}$ and compute $\tilde \K$
as:
\begin{align} \label{Eqn: tilde U, def}
    \tilde{\boldU} &:= \begin{bmatrix}
        \tilde{\boldu}^\paren{2} - \tilde{\boldu}^\paren{1} & \cdots & \tilde{\boldu}^\paren{D} - \tilde{\boldu}^\paren{D-1}
    \end{bmatrix}, \\ 
    \label{Eqn: tilde Y, def}
    \tilde{\boldY} &:= \begin{bmatrix}
        \tilde{\boldy}^\paren{2} - \tilde{\boldy}^\paren{1} & \cdots & \tilde{\boldy}^\paren{D} - \tilde{\boldy}^\paren{D-1}
    \end{bmatrix},
    \\
    \label{Eqn: K from U and Y matrices}
    \tilde \K &:= \tilde{\boldU} \tilde{\boldY}^\dagger,
\end{align}
\revisionRR{where $\dagger$ denotes the Moore-Penrose pseudoinverse.}
We then compute the associated system response $\tilde \Phi$ via \eqref{Eqn: Phi from K}.

Next, to infer the nominal trajectories $(\boldz^\star, \boldv^\star)$, we stack the equalities $z_{t+1}^\star = A_t z_t^\star + B_t v_t^\star$ across $t \in [T]$ and rearrange \eqref{Eqn: Demonstration Generation, Control Design via Output Feedback, Stacked} to obtain that, for each $d \in [D]$:
\begin{align} \nonumber
    (I - \Z \A) \boldz^\star - \Z \B \boldv^\star &= \boldzero, \\ \nonumber
    -\K^\star \C \boldz^\star + \boldv^\star &= (\tilde{\boldu}^\paren{d} - \K^\star \tilde{\boldy}^\paren{d}) - (\bolddelta_u^\paren{d} - \K^\star \bolddelta_y^\paren{d}).
\end{align}
We then estimate $(\boldz^\star, \boldv^\star)$ from $\tilde \K$ as follows\footnote{For the case of nonlinear dynamics, see Remark \ref{Remark: Nonlinear dynamics, nominal trajectory and control recovery via nonlinear regression}.}:
\begin{align} \label{Eqn: z, v from tilde K, tilde D}
    \begin{bmatrix}
        \tilde{\boldz} \\ \tilde{\boldv}
    \end{bmatrix} 
    &= \frac{1}{d} \sum_{d=1}^D
    \begin{bmatrix}
        I - \Z\A & -\Z\B \\ - \tilde \K \C & I
    \end{bmatrix}^\dagger 
    \begin{bmatrix}
        \boldzero \\ \tilde{\boldu}^\paren{d} - \tilde \K \tilde{\boldy}^\paren{d}
    \end{bmatrix}.
\end{align}
Above, we average over the terms $\{\tilde{\boldu}^\paren{d} - \tilde \K \tilde{\boldy}^\paren{d}: d \in [D] \}$ to reduce the impact of the transmission errors $\{\bolddelta_u^\paren{d}, \bolddelta_y^\paren{d}: d \in [D] \}$ on the recovery of $(\boldz^\star, \boldv^\star)$. 
Finally, we infer the unknown constraint parameter $\theta^\star$. To ease notation, given $\boldz, \boldv, \Phi$, we define $\boldeta := (\boldz, \boldv, \Phi)$.
We also define \revisionB{$\boldnu$}, $\boldlambda_k$, and $\boldlambda_\urk$ to denote the Lagrange multipliers for \eqref{Eqn: Reformulated, Robust Forward, Equality Constraints}, \eqref{Eqn: Reformulated, Robust Forward, Inequality Constraints, Known}, and \eqref{Eqn: Reformulated, Robust Forward, Inequality Constraints, Unknown}, respectively.
The KKT conditions of \eqref{Eqn: Reformulated, Robust Forward} are then:
\begin{subequations} \label{Eqn: KKT, Forward Game}
\begin{align} \label{Eqn: KKT, Forward, Primal Feasibility}
    &\boldh(\boldeta) = \revisionRR{\boldzero}, \hspace{5mm} \boldg_k(\boldeta) \leq \boldzero, \hspace{5mm} \boldg_\urk(\boldeta, \theta) \leq \revisionRR{\boldzero}, \\ \label{Eqn: KKT, Forward, Lagrange Multiplier non-negativity}
    &\boldlambda_k, \boldlambda_\urk \geq \revisionRR{\boldzero}, \\ \label{Eqn: KKT, Forward, Complementary Slackness}
    &\boldlambda_k \odot \boldg_k(\boldeta) = \boldzero, \hspace{5mm} \boldlambda_\urk \odot \boldg_\urk(\boldeta, \theta) = \revisionRR{\boldzero}, \\ \label{Eqn: KKT, Forward, Stationarity}
    &\nabla_{\boldeta} J(\boldeta) + \boldlambda_k^\top \nabla_{\boldeta} \boldg_k(\boldeta) \\ \nonumber
    & \hspace{2mm} + \boldlambda_\urk^\top \nabla_{\boldeta} \boldg_\urk(\boldeta, \theta) 
    + \boldnu^\top \nabla_{\boldeta} \boldh(\boldeta) = \revisionRR{\boldzero^\top},
\end{align}
\end{subequations}
\revisionRR{where $\odot$ denotes element-wise multiplication.}
We denote by $\KKT(\boldeta)$ the set of all $(\theta, \boldlambda_k, \boldlambda_\urk, \boldnu)$ satisfying \eqref{Eqn: KKT, Forward Game}.
Since the true nominal trajectory and output response computed by the demonstrator, which we denote by $\boldeta^\star := (\boldz^\star, \boldv^\star, \Phi^\star)$, minimizes \eqref{Eqn: Reformulated, Robust Forward}, there exist Lagrange multiplier values $\boldlambda_k^\star, \boldlambda_\urk^\star, \boldnu^\star$ such that $(\theta^\star, \boldlambda_k^\star, \boldlambda_\urk^\star, \boldnu^\star)$ solves the following feasibility problem with $\boldeta = \boldeta^\star$:
\begin{subequations}
\label{Eqn: KKT, Inverse, Optimal, Feasibility Problem}
\begin{align} \label{Eqn: KKT, Inverse, Optimal, Objective}
    \text{find} \hspace{5mm} &\theta, \boldlambda_k, \boldlambda_\urk, \boldnu, \\ \label{Eqn: KKT, Inverse, Optimal, Constraints}
    \text{s.t.} \hspace{5mm} &(\theta, \boldlambda_k, \boldlambda_\urk, \boldnu) \in \KKT(\boldeta).
\end{align}
\end{subequations}
Conversely, solutions to \eqref{Eqn: KKT, Inverse, Optimal, Feasibility Problem} with $\boldeta = \boldeta^\star$ describe parameter values 
consistent with the optimality of $\boldeta^\star$ with respect to \eqref{Eqn: Reformulated, Robust Forward}\revisionRR{; the ground truth $\theta^\star$ is one such parameter value.} 
Concretely,
given any $\boldeta = (\boldz, \boldv, \Phi)$, we denote by $\F(\boldeta) \subseteq \R^d$ the set of $\KKT$-compatible parameter values, i.e., 
\begin{align} \label{Eqn: F(eta)}
    \F(\boldeta) &:= \{\theta \in \R^d: \exists \ \boldlambda_k, \boldlambda_\urk, \boldnu \\ \nonumber
    &\hspace{2cm} \st (\theta, \boldlambda_k, \boldlambda_\urk, \boldnu) \in \KKT(\boldeta). \}
\end{align}

\subsection{Learning Guarantees Under Zero Transmission Error}
\label{subsec: Constraint Learning Guarantees Under Zero Transmission Error}

\revisionRR{First, we prove that the following matrix $\Gamma$ is invertible:
\begin{align} \label{Eqn: Block matrix for recovering nominal trajectories, def}
\Gamma :=
\begin{bmatrix}
    I - \Z\A & -\Z\B \\ -\K^\star \C & I
\end{bmatrix} 
\end{align}
}
\begin{lemma}
When $f_t(\cdot)$ is LTV,
$\Gamma$
is invertible.
\end{lemma}
\begin{proof}
By the Schur complement, $\det(\Gamma) = \det(I - \Z\A) \det(I - \Z(\A + \B\K^\star\C))$. Since $\A$, $\B$, $\K^\star$, $\C$ are lower block triangular and $\Z$ is the downshift operator, $\Z\A$ and $\Z(\A + \B\K^\star\C)$ are both \textit{strictly} lower block triangular. Thus, $\det(\Gamma) = 1$, so $\Gamma$ is invertible.
\end{proof}

We present a condition under which $\boldeta^\star$, the true nominal trajectories and system response, can be accurately recovered.

\begin{assumption} \label{Assump: Full Rank}
$\tilde{\boldY}$ has full row rank.
\end{assumption}

In words, Assumption \ref{Assump: Full Rank} states that the 
demonstrations $\tilde{\boldY}$ are sufficiently rich for recovering $\K^\star$, 
and that $\K^\star$ satisfies a mild regularity condition. In particular, 
\revisionRR{if 
$\{\bolddelta_u^\paren{d}: d \in [D]\}$, $\{\bolddelta_y^\paren{d}: d \in [D]\}$ 
are i.i.d. and independent of $\{w_t^\paren{d}, e_t^\paren{d}:t \in [T], d \in [D] \}$},
then $\tilde{\boldY}$ has full row rank 
whenever 
$D \geq n_o (T+1) + 1$.
Thm. \ref{Thm: Recovering eta star} below shows that under zero transmission noise, Assumption \ref{Assump: Full Rank} is sufficient to guarantee the accurate recovery of $\boldeta^\star$.

\begin{theorem} \label{Thm: Recovering eta star}
If \revisionRR{$f_t(\cdot)$ is LTV}, Assumption \ref{Assump: Full Rank} holds, and $(\bolddelta_u^\paren{d}, \bolddelta_y^\paren{d}) = \boldzero, \forall d \in [D]$, 
then $\tilde \boldeta = (\tilde{\boldz}, \tilde{\boldv}, \tilde \Phi)$, as computed by \eqref{Eqn: K from U and Y matrices}, \eqref{Eqn: Phi from K}, and \eqref{Eqn: z, v from tilde K, tilde D}, equals $\boldeta^\star = (\boldz^\star, \boldv^\star, \Phi^\star)$.
\end{theorem}

\begin{proof}
When $(\bolddelta_u^\paren{d}, \bolddelta_y^\paren{d}) = \boldzero, \forall d \in [D]$, \eqref{Eqn: tilde u, K star, tilde y, affine equality} becomes:
\begin{align}
    \tilde{\boldu}^\paren{d} - \tilde{\boldu}^\paren{d-1} &= \K^\star (\tilde{\boldy}^\paren{d} - \tilde{\boldy}^\paren{d-1}),
\end{align}
which can be stacked across all $d \in [2, D]$ as $\tilde{\boldU} = \K^\star \tilde{\boldY}$. Since $\tilde{\boldY}$ has full row rank, $\tilde{\boldY} \tilde{\boldY}^\dagger = I_{n_o \revisionRR{(T+1)}}$, so $\K^\star = \tilde{\boldU} \tilde{\boldY}^\dagger$. Together with \eqref{Eqn: K from U and Y matrices} and \eqref{Eqn: Phi from K}, this confirms that $\K^\star = \tilde \K$ and $\Phi^\star = \tilde \Phi$. Then, from \eqref{Eqn: z, v from tilde K, tilde D} and preceding equations, we have under Assumption \ref{Assump: Full Rank} that
$(\tilde{\boldz}, \tilde{\boldv}) = (\boldz^\star, \boldv^\star)$. 
\end{proof}

\begin{remark} \label{Remark: Nonlinear dynamics, nominal trajectory and control recovery via nonlinear regression}
If the dynamics \eqref{Eqn: Forward Dynamics} are nonlinear, we can still recover the output feedback law via \eqref{Eqn: K from U and Y matrices} (since the output map is assumed LTV). However, we would apply nonlinear regression techniques to fit the input-output demonstrations, while enforcing the dynamics $z_{t+1}^\star = f_t(z_t^\star, v_t^\star)$, instead of applying the procedure described in Sec. \ref{subsec: Recovering the Nominal Trajectory, Output Feedback, and Constraint Parameters}. 
Although the nominal trajectories would no longer be guaranteed to be accurately recovered (in the sense of Thm. \ref{Thm: Recovering eta star}),
we find that in simulation, we still recover the true nominal trajectories with minimal error when enough demonstrations are provided.
\end{remark}



We next define the 
set of \textit{guaranteed safe} (resp., \textit{unsafe}) nominal trajectories and output responses, denoted $\G_s(\boldeta^\star)$ (resp., $\G_\urs(\boldeta^\star)$), to consist of nominal trajectories and output responses
which are safe (resp., unsafe) with respect to all 
$\theta \in \F(\boldeta^\star)$.
The set of guaranteed safe nominal trajectories and output feedback laws can then be used to generate motion plans downstream that will remain safe despite even worst-case noise realizations.
Concretely, define:\footnote{
We remark that 
$\G_\urs(\boldeta^\star) \ne \G_s(\boldeta^\star)^c$, i.e., a given nominal trajectory and system response may be neither guaranteed safe nor guaranteed unsafe.}
\begin{align} 
\nonumber
    \G_s(\boldeta^\star) &:= \textstyle\bigcap_{\theta \in \F(\boldeta^\star)} \big\{ \boldeta': \boldg(\boldeta', \theta) \leq 0 \big\} = \bigcap_{\theta \in \F(\boldeta^\star)} \revisionRR{\Safe}(\theta), \\  
    \nonumber
    \G_\urs(\boldeta^\star) &:= \textstyle\bigcap_{\theta \in \F(\boldeta^\star)} \big\{ \boldeta': \boldg(\boldeta', \theta) \leq 0 \big\}^c = \bigcap_{\theta \in \F(\boldeta^\star)} \revisionRR{\Avoid}(\theta).
\end{align}

Below, Thm. \ref{Thm: Conservativeness of Safe and Unsafe Set Recovery from KKT, Inverse, Optimal}
states that $\G_s(\boldeta^\star)$ is guaranteed to inner-approximate (i.e., conservatively estimate) $\revisionRR{\Safe}(\theta^\star)$, the set of nominal trajectories and system responses $\boldeta$ that are safe with respect to the ground truth inequality constraints.
Similarly, $\G_\urs(\boldeta^\star)$ always inner-approximates $\revisionRR{\Avoid}(\theta^\star)$.

\begin{theorem}
\label{Thm: Conservativeness of Safe and Unsafe Set Recovery from KKT, Inverse, Optimal}
$\G_s(\boldeta^\star) \subseteq \revisionRR{\Safe}(\theta^\star)$ and $\G_\urs(\boldeta^\star) \subseteq \revisionRR{\Avoid}(\theta^\star)$.
\end{theorem}

\begin{proof}
By the definitions of $\G_s(\boldeta^\star)$ and $\G_\urs(\boldeta^\star)$, it suffices to prove $\theta^\star \in \F(\boldeta^\star)$. Since $\boldeta^\star$ solves \eqref{Eqn: Reformulated, Robust Forward}, there exist Lagrange multipliers $\boldlambda_k, \boldlambda_\urk, \boldnu$ such that $(\theta^\star, \boldlambda_k, \boldlambda_\urk, \boldnu) \in \KKT(\boldeta^\star)$. Thus, by definition of $\F(\boldeta^\star)$, we have $\theta^\star \in \F(\boldeta^\star)$,
so $\G_s(\boldeta^\star) \subseteq \revisionRR{\Safe}(\theta^\star)$ and $\G_\urs(\boldeta^\star) \subseteq \revisionRR{\Avoid}(\theta^\star)$.
\end{proof}


\subsection{Sensitivity Analysis With Respect to Transmission Error}
\label{subsec: Sensitivity Analysis with respect to Transmission Error}

We now prove that when the transmission noise $\{(\bolddelta_u^\paren{d}, \bolddelta_y^\paren{d}): d \in [D] \}$ is nonzero, 
\revisionRR{the error in the recovered output feedback $\K$ 
and nominal trajectory $(\boldz, \boldv)$ 
scales at most linearly in the noise magnitudes $\Vert \bolddelta_u^\paren{d} \Vert_2$ and $\Vert \bolddelta_y^\paren{d} \Vert_2$.}



\begin{theorem} \label{Thm: Sensitivity Analysis w.r.t. Transmission Error}
Suppose \revisionRR{$f_t(\cdot)$ is LTV}, Assumption \ref{Assump: Full Rank} holds, and $\max_{d \in [D]} \max\{ \Vert \bolddelta_u^\paren{d} \Vert_2, \Vert \bolddelta_y^\paren{d} \Vert_2\}  < \epsilon$ for some $\epsilon > 0$. \revisionRR{Suppose} $\boldY :=
\begin{bmatrix}
    \boldy^\paren{2} - \boldy^\paren{1} & \cdots & \boldy^\paren{D} - \boldy^\paren{D-1}
\end{bmatrix}$ \revisionRR{has full row rank}, and set
$\rho_1 := \sqrt{D-1} \Vert \boldY^\dagger \Vert_2$,
\revisionRR{$\rho_2 := \sqrt{D-1} \Vert \boldY^\dagger \Vert_2 \Vert \Gamma^{-1} \Vert_2 \Vert \C \Vert_2$},
\revisionRR{$\rho_3 := \sqrt{D-1} \Vert \boldY^\dagger \Vert_2 \Vert \Gamma^{-1} \Vert_2$}, and 
\revisionRR{$\rho_4 := \Vert \Gamma^{-1} \Vert_2 \big( \Vert \K^\star \Vert_2 + 1 \big).$}
Let $(\delta \boldz, \delta \boldv, \revisionB{\delta \K}) := (\tilde{\boldz} - \boldz^\star, \tilde{\boldv} - \boldv^\star, \tilde \K - \K^\star)$. Then:
\begin{subequations} \label{Eqn: delta K, delta z, delta v bounds}
\begin{align} 
\nonumber
    \Vert \delta \K \Vert_2 &\leq \rho_1 \big( \Vert \tilde \K \Vert_2 + 1 \big) \epsilon, \\ 
    \nonumber
    \left\Vert \begin{bmatrix}
        \delta \boldz \\ \delta \boldv
    \end{bmatrix} \right\Vert_2 &\leq \left( \rho_2 \left\Vert \begin{bmatrix}
        \tilde{\boldz} \\ \tilde{\boldv}
    \end{bmatrix} \right\Vert_2 + \rho_3 \Vert \tilde{\boldy}^\paren{1} \Vert_2 \right) (\Vert \tilde \K \Vert_2 + 1) \epsilon + \rho_4 \epsilon.
\end{align}
\end{subequations}
\end{theorem}

\begin{proof}
We apply least-squares perturbation analysis
\cite[Sec 2.2, \revisionRR{Eqn. (2.2)}]{Demmel1997AppliedNumericalLinearAlgebra}
to analogs of \eqref{Eqn: K from U and Y matrices} and \eqref{Eqn: z, v from tilde K, tilde D} that relate $(\tilde{\boldz}, \tilde{\boldv}, \tilde \K)$ to $\{ (\tilde{\boldu}^\paren{d}, \tilde{\boldy}^\paren{d}): d \in [D] \}$ and $(\boldz^\star, \boldv^\star, \K^\star)$ to $\{ (\boldu^\paren{d}, \boldy^\paren{d}): d \in [D] \}$, respectively. 
\revisionRR{Concretely, from \cite{Demmel1997AppliedNumericalLinearAlgebra}, if $MX = P$ and $(M + \delta M) \hat X = P + \delta P$ for dimension-compatible matrices $M, P, X, \delta M, \delta P, \hat X$,
and $M$ has full column rank,
then:
$\Vert \delta X \Vert_2 \leq \Vert M^\dagger \Vert_2 (\Vert \delta M \Vert_2 \cdot \Vert \hat X \Vert_2 + \Vert \delta P \Vert_2).$}
Details are omitted for brevity.
\end{proof}

\section{Experiments}
\label{sec: Experiments}

To evaluate our method, we perform constraint learning on simulated 4D double integrator, 4D unicycle, nonlinear 6D quadcopter, and linearized 12D quadcopter dynamics 
\cite{sabatino2015PhDThesisquadrotor}. We use Gurobi 
to solve all optimization problems.
\revisionRR{
Our code and additional results can be found at \href{https://github.com/zhangzdd/SLS-ConstraintLearning}{https://github.com/zhangzdd/SLS-ConstraintLearning.}
}



\subsection{Experiment Setup}
\label{subsec: Experiment Setup}

We encode polytopic collision-avoidance constraints using intersections of unions of half-spaces, as shown below:
\begin{align*}
    \revisionRR{S_\urk}(\theta) &= 
    \wedge_{\alpha \in [\overline N_c]} \vee_{\beta \in [N_c]} \big\{ A_{\alpha, \beta}(\theta) (\boldx, \boldu) \leq b_{\alpha, \beta}(\theta) \big\},
\end{align*}
where $A_{\alpha, \beta}(\theta)$ and \revisionB{$b_{\alpha, \beta}(\theta)$} are dimension-compatible $\forall \alpha, \beta$. 

To define agent costs $J$, given a state trajectory $\boldx \in \R^{nT}$, let the vector $p_t$ and the scalars $p_{x,t}, p_{y,t}$ denote components of $\boldx$ describing the position vector, $x$-coordinate and $y$-coordinate at time $t$, 
respectively. We then define:
\begin{subequations}
\begin{align} \label{Eqn: Exp, J1}
    J^\paren{1}(\boldx) &= \textstyle\sum_{t=0}^{T-1} \big( \Vert p_{t+1} - p_t \Vert_2^2 - p_{x,t} \big), \\ \label{Eqn: Exp, J2}
    J^\paren{2}(\boldx) &= \textstyle\sum_{t=0}^{T-1} \big( \Vert p_{t+1} - p_t \Vert_2^2 - p_{x,t} - p_{y,t} \big), \\ \label{Eqn: Exp, J3}
    J^\paren{3}(\boldx) &= \textstyle\sum_{t=0}^{T-1} \Big( \Vert p_{t+1} - p_t \Vert_2^2 + \frac{1}{T} \Vert p_t - p_T \Vert_2^2 \Big).
\end{align}
\end{subequations}
Demonstrations in Figs. \ref{fig: SLS}c and \ref{fig: SLS}d-f were generated with $J^\paren{3}$ and $J^\paren{2}$, respectively; all others were generated using $J^\paren{1}$.

\begin{figure}
      \begin{minipage}[c]{0.5\linewidth}
        \includegraphics[width=0.99\linewidth]{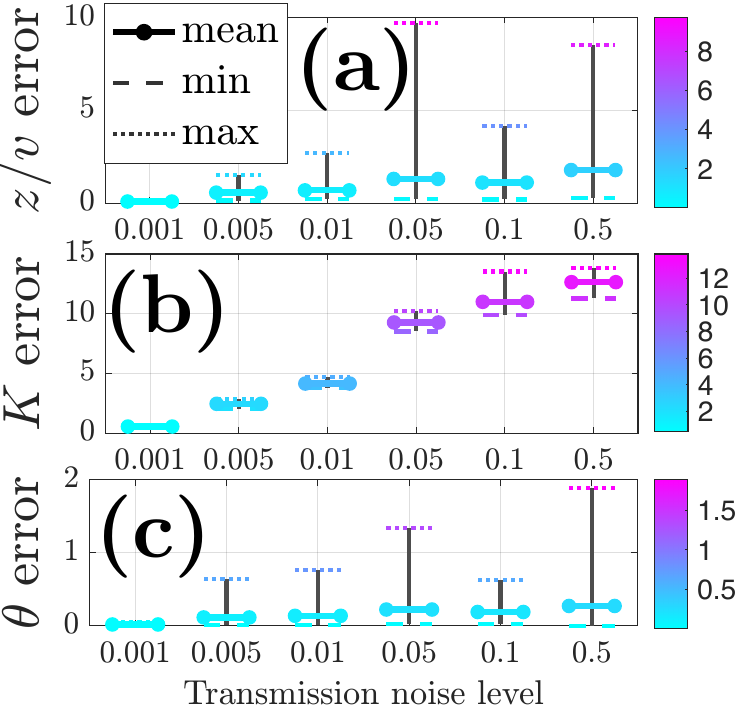}
      \end{minipage}
      \begin{minipage}[c]{0.49\linewidth}
            \caption{
            Error magnitude in recovered (a) nominal trajectories and
            controls, (b) output feedback, and (c) constraint parameter as a function of the transmission error level. 
            Demonstrations were generated using noise-corrupted double integrator dynamics and (full) state feedback SLS controllers.
            On average, our method learns \textit{a priori} unknown parameters 
            with high accuracy 
            under low levels of transmission noise. 
    }
        \label{fig: Transmission Error}
      \end{minipage}
\end{figure}

In simulations for 
\revisionRR{Figs. \ref{fig: SLS}c, \ref{fig: CCM}a, \ref{fig: CCM}b, \ref{fig: PD}a, \ref{fig: PD}b, \ref{fig: PD}c}, 
the set $W$ bounding the 
terms $w_t$ was respectively defined to be $\infty$-norm balls of radii \revisionRR{0.05, 0.05, 0.05, 0.2, 0.07, and 0.05}; in all other simulations, $W$ was set as the unit $\infty$-norm ball. For Fig. \ref{fig: SLS}, the set $V$ bounding the \revisionRR{output} noise terms \revisionRR{$e_t$} was set as the norm ball of radius 0.02; all other simulations used zero output noise. As shown in Figs. \ref{fig: SLS}-\ref{fig: PD}, for each experiment,
we fix a ground truth constraint set 
\revisionRR{(shaded gray)},
and use output SLS, 
control contraction metrics (CCM), 
or proportional–derivative (PD) control
to generate 
nominal trajectories (green, with circles/stars for origins/goals), output feedback laws, and demonstration rollouts (blue) with corresponding 
\revisionRR{state trajectory bounds (edges in magenta)}\footnote{\revisionRR{Such bounds are computed using $\Phi(\boldw, \bolde)$, the maximum deviation from the nominal, and realized closed-loop, trajectories
(Sec. \ref{subsec: System Level Synthesis (SLS) for Output Feedback}).}}.
We then apply our methods (Sec. \ref{sec: Methods}) to learn nominal trajectories, output feedback laws, 
and constraints (boundaries in black) from the given demonstrations. 
For each experiment, we generate demonstrations either via (full) state feedback (i.e., $y_t = x_t$), or output feedback, wherein only the position vector is observed (i.e., $y_t = p_t$), and we set \revision{$D = 100$}.
The simulations in Figs. \ref{fig: SLS}-\ref{fig: PD} are generated with zero transmission noise; Fig. \ref{fig: Transmission Error} shows that the learning accuracy of our method degrades gracefully as transmission error magnitude increases.

\subsection{Simulation Results}
\label{subsec: Simulation Results}

\begin{figure}
    \centering
    \includegraphics[width=0.99\linewidth]{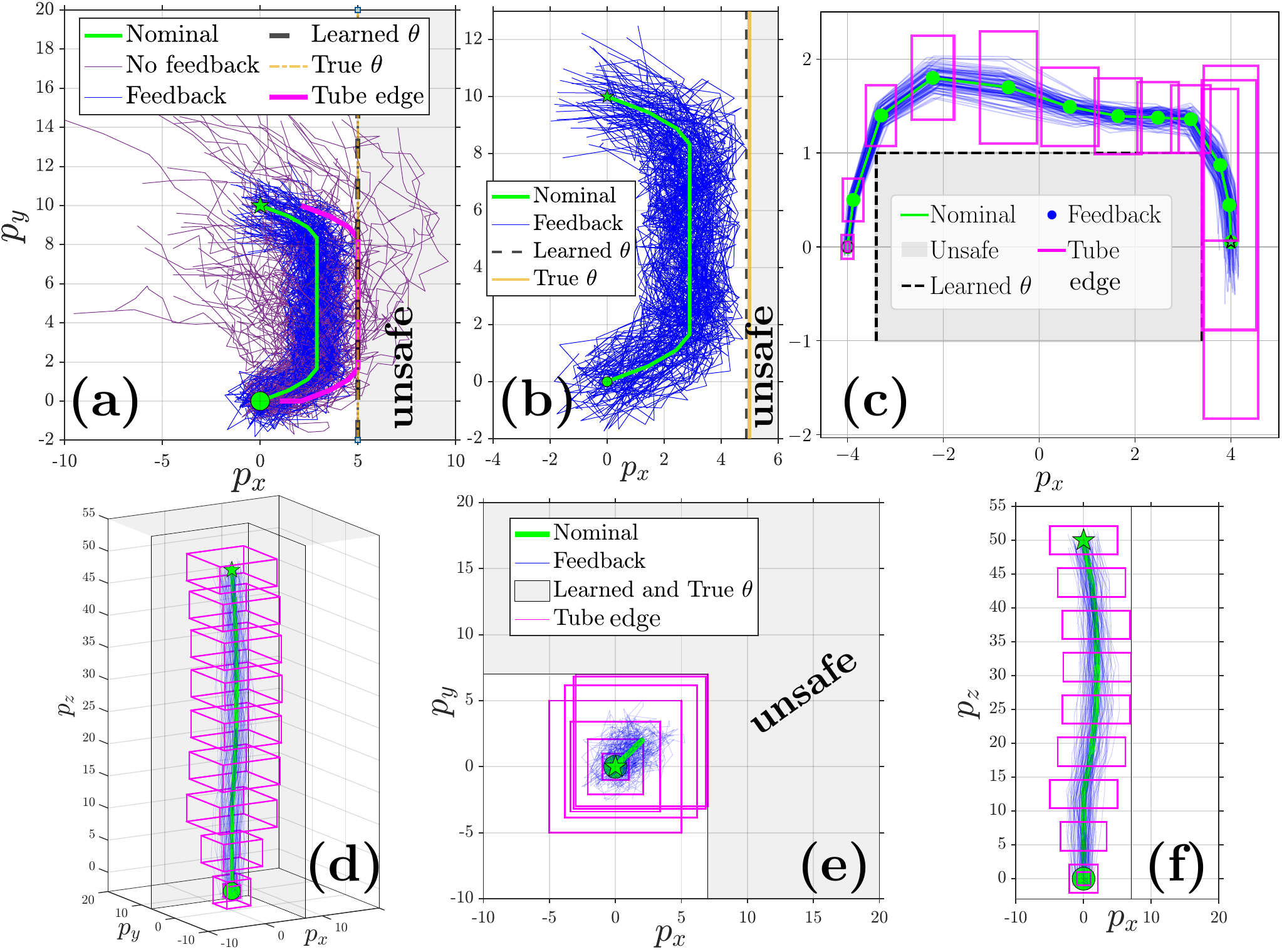}
    \caption{
    Constraint learning from demonstrations generated via SLS with double integrator 
    (a, c) 
    and linear 12D quadcopter 
    (d, e, f)
    dynamics. (e, f) provide views of (d) from different angles. (a, c-f) Our method accurately learns true collision avoidance constraints across all simulations,
    but (b) the baseline \cite{Chou2020LearningConstraintsFromLocallyOptimalDemonstrationsUnderCostFunctionUncertainty} did not;
    note the mismatch between the learned (black) and true (yellow) constraints.
    }
    \label{fig: SLS}
\end{figure}

\textit{a) SLS Controllers}:
We present simulations evaluating our method (Figs. \ref{fig: SLS}a, c-f) and comparing it to the baseline 
in \cite{Chou2020LearningConstraintsFromLocallyOptimalDemonstrationsUnderCostFunctionUncertainty} (Fig. \ref{fig: SLS}b), which was designed to learn constraints from noise-free, fully observed state trajectories. 
We
learn constraints from demonstrations generated by unrolling noise-corrupted double integrator (Fig. \ref{fig: SLS} a-c) or linearized 12D quadcopter (Fig. \ref{fig: SLS}d-f) dynamics, and applying a state feedback 
(Fig. \ref{fig: SLS} a-b, d-f) or output feedback (Fig. \ref{fig: SLS}c) SLS controller 
(Fig. \ref{fig: SLS}c). 
The nominal trajectories and feedback laws underlying the demonstration were jointly generated via \eqref{Eqn: Reformulated, Robust Forward}. 
Across experiments in Figs. \ref{fig: SLS}a, c-f, our methods, as described in \eqref{Eqn: K from U and Y matrices}-\eqref{Eqn: KKT, Inverse, Optimal, Feasibility Problem}, accurately learned the true nominal trajectories, output feedback laws, and constraint parameters underlying the input-output demonstrations. In contrast, the baseline \cite{Chou2020LearningConstraintsFromLocallyOptimalDemonstrationsUnderCostFunctionUncertainty} fails to learn the true underlying constraints (Fig. \ref{fig: SLS}b). 


\begin{figure}
      \begin{minipage}[c]{0.69\linewidth}
        \includegraphics[width=0.99\linewidth]{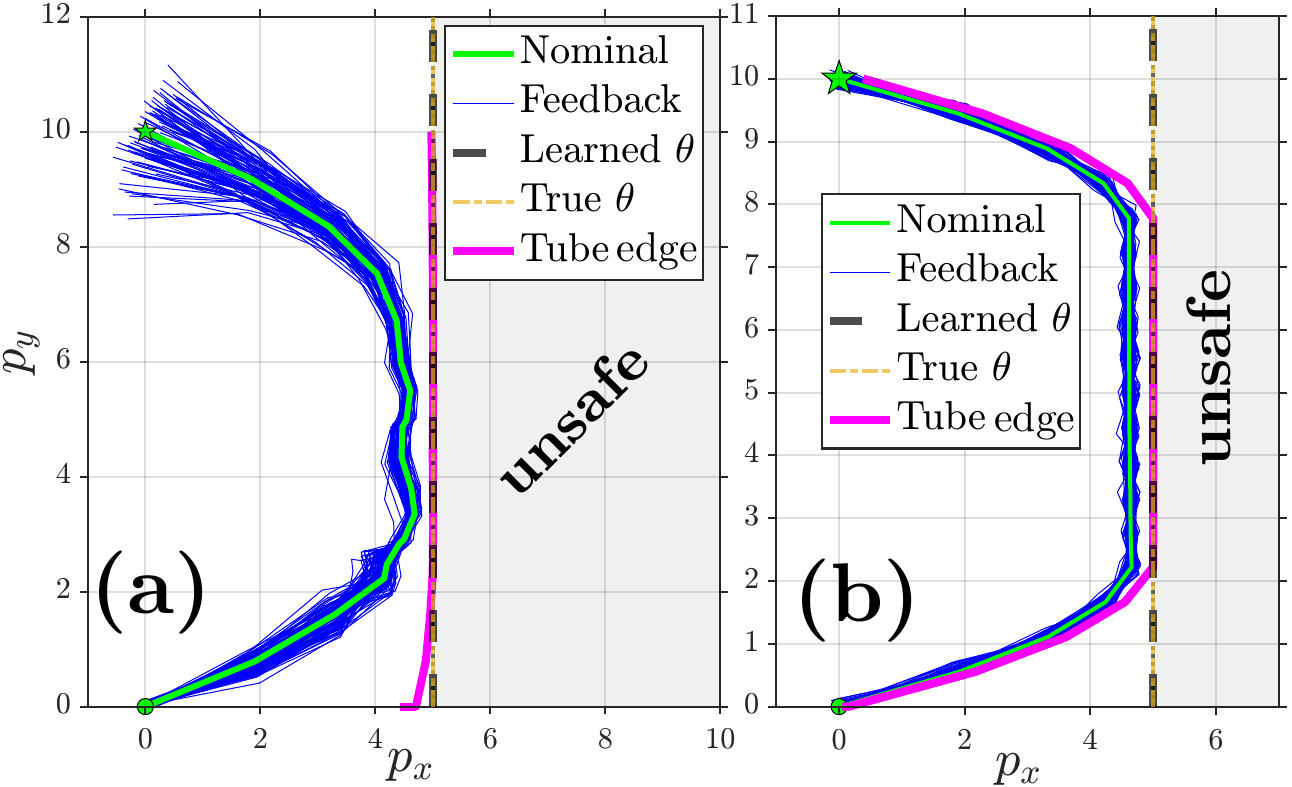}
      \end{minipage}
      \begin{minipage}[c]{0.3\linewidth}
            \caption{
            Constraint learning from demonstrations produced via CCM with unicycle (a) and double integrator (b) dynamics. Our method accurately learns 
            true 
            collision avoidance
            constraints across all simulations. 
    }
        \label{fig: CCM}
      \end{minipage}
\end{figure}

\textit{b) Control Contraction Metrics (CCM)-Based State Feedback Controllers}:
We 
learn constraints from
demonstrations generated by unrolling noise-corrupted unicycle 
(Fig. \ref{fig: CCM}a)
or double integrator 
(Fig. \ref{fig: CCM}b)
dynamics with full state observation, using the CCM-based state feedback law in \cite{ManchesterSlotine2017ControlContractionMetrics}. 
Concretely, we computed the error feedback law $\K$ of \cite[Sec. III-B]{ManchesterSlotine2017ControlContractionMetrics}, and the associated system response $\Phi$ via \eqref{Eqn: Phi from K}. We then use \eqref{Eqn: Reformulated, Robust Forward} to compute the nominal trajectory for generating demonstrations, while fixing $\Phi$ fixed at the value derived from 
via CCM
(see Remark \ref{Remark: Nonlinear dynamics, Robust reformulation}).
By applying \eqref{Eqn: K from U and Y matrices}-\eqref{Eqn: KKT, Inverse, Optimal, Feasibility Problem}, we accurately learned the true nominal trajectories, output feedback laws, and constraint parameters (Fig. \ref{fig: CCM}). 


\begin{figure}
    \centering
    \includegraphics[width=0.99\linewidth]{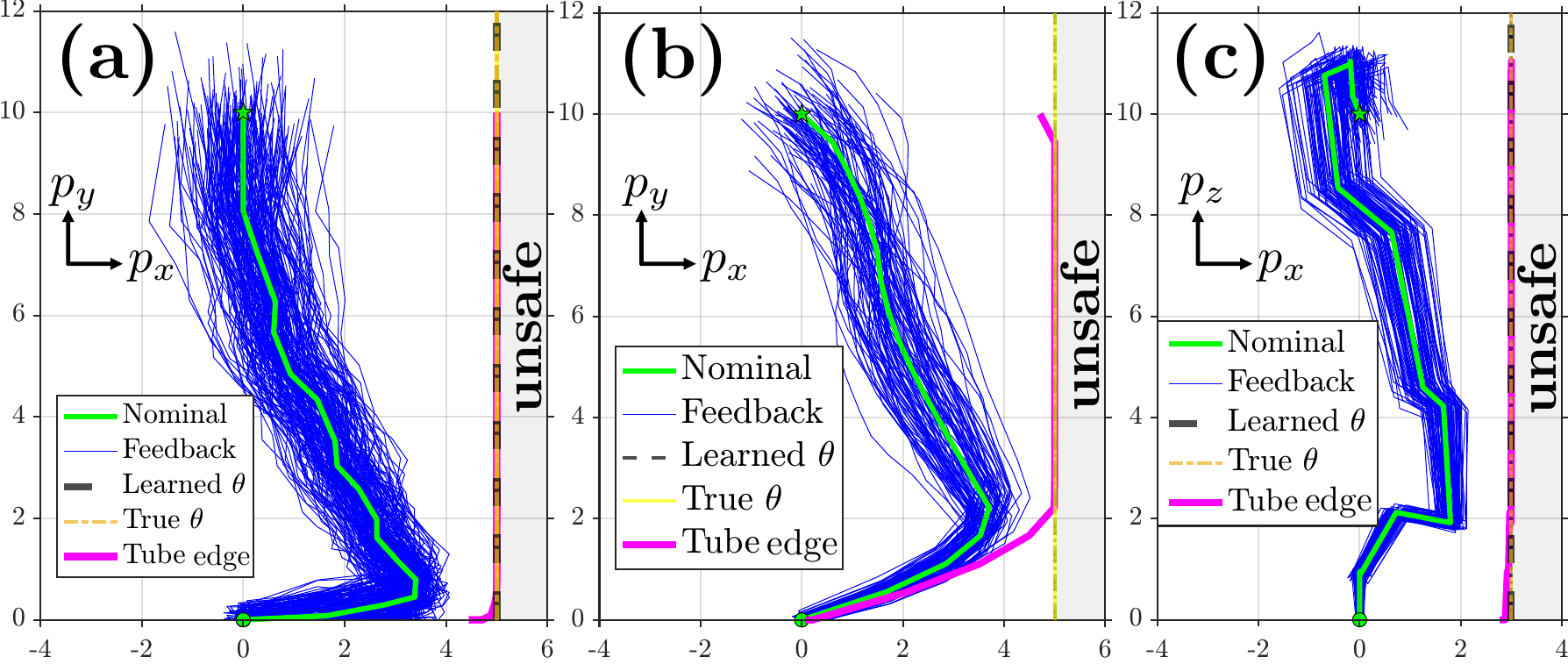}
    \caption{
    Constraint learning from demonstrations produced via \revisionRR{PD} with (a) unicycle, (b) double integrator, and (c) nonlinear 6D quadcopter dynamics. Our method accurately learns 
    true 
    collision avoidance 
    constraints for all simulations. 
    }
    \label{fig: PD}
\end{figure}

\textit{c) Proportional-Derivative (PD) Controllers}:
We recover constraints from demonstrations generated using noise-corrupted unicycle (Fig. \ref{fig: PD}a), double integrator (Fig. \ref{fig: PD}b), or nonlinear 6D quadcopter (Fig. \ref{fig: PD}c) dynamics, by applying PD state feedback controllers. 
Similar to the CCM-based methods, we first compute the feedback policy $\K$ and the corresponding system response $\Phi$ via PD control techniques and \eqref{Eqn: Phi from K}. 
To obtain a nominal trajectory from which 
demonstrations are then generated,
we solve \eqref{Eqn: Reformulated, Robust Forward} while holding $\Phi$ fixed at its value computed via the PD control method
and \eqref{Eqn: Phi from K}.
In the Fig. \ref{fig: PD} experiments, our methods, as described by \eqref{Eqn: K from U and Y matrices}-\eqref{Eqn: KKT, Inverse, Optimal, Feasibility Problem} accurately recovered the true nominal trajectories, output feedback laws, and constraint parameters that characterize the input-output demonstrations.

\section{Conclusion}
\label{sec: Conclusion and Future Work}

We presented a novel IOC-based method for recovering unknown parametric constraints from locally optimal input-output demonstrations, generated by unrolling stochastic dynamics while applying an output feedback law designed using robust optimal control. We present theory and simulations to demonstrate that our method can accurately recover the demonstrator's
output feedback laws and constraints.

\renewcommand*{\bibfont}{\footnotesize}
\printbibliography


\end{document}